\documentclass[final,5p,times,twocolumn,authoryear]{elsarticle}

\usepackage{amsmath,amssymb,mathtools}
\usepackage{xcolor}
\usepackage{booktabs,longtable}
\usepackage{subcaption}
\usepackage{siunitx}
\usepackage{hyperref}
\usepackage{cuted}
\usepackage{orcidlink}

\definecolor{darkgreen}{RGB}{0,128,0}

\journal{Nuclear Physics A}

\begin{document}
\begin{frontmatter}

\title{Half-Life Measurements of $^{110}$Sn, $^{113}$Sn, $^{117\mathrm{m}}$Sn, and $^{123\mathrm{m}}$Sn Produced via Photon Activation of Natural Tin}
\author[ns]{O.~Nusair\,\orcidlink{0000-0002-4841-5286}\corref{cor1}}
\ead{omar_nusair@yahoo.com}
\author[ns]{D.~DeVries}
\author{M.~Toro-Gonzalez\,\orcidlink{0000-0003-3320-423X}}

\cortext[cor1]{Corresponding author}

\affiliation[ns]{organization={NorthStar Medical Radioisotopes, LLC},
                 city={Beloit},
                 state={WI},
                 postcode={53511},
                 country={USA}}

%% ---------------- Abstract ----------------
\begin{abstract}
We report independent determinations of the ground-state half-lives of $^{110}$Sn, $^{113}$Sn, and the isomeric states $^{117\mathrm{m}}$Sn (J$^{\pi} = 11/2^{-}$) and $^{123\mathrm{m}}$Sn (J$^{\pi} = 3/2^{+}$), produced via photon activation of natural tin using a TT-300HE Rhodotron accelerator. The activated samples were monitored over several months using a high-purity germanium (HPGe) detector. Time-dependent $\gamma$-ray spectra were analyzed using Gaussian peak fitting for the \SI{280.49}{keV}, \SI{391.697}{keV}, \SI{158.56}{keV}, and \SI{160.34}{keV} transitions, yielding half-lives of \SI{4.165(25)}{h} for $^{110}$Sn, \SI{116.08(94)}{d} for $^{113}$Sn, \SI{13.95(1)}{d} for $^{117\mathrm{m}}$Sn, and \SI{39.95(12)}{min} for $^{123\mathrm{m}}$Sn. Agreement with Nuclear Data Sheets (NDS) recommended values is generally observed for $^{110}$Sn, $^{113}$Sn, and $^{123\mathrm{m}}$Sn, with deviations consistent within combined uncertainties when quantified using standardized differences (z-scores). In contrast, $^{117\mathrm{m}}$Sn exhibits a statistically significant deviation from the evaluated value of \SI{13.76(4)}{d}, with a z-score indicating a discrepancy well beyond expected statistical fluctuations. This result suggests a systematic difference warranting further investigation, with potential implications for applications relying on precise decay data, including calibration, dosimetry, and astrophysical modeling.
\end{abstract}


\end{frontmatter}

%% ---------------- Main text ----------------

\section{Introduction}

Precise radionuclide half-lives are fundamental to applied and basic nuclear science. They underpin normalization of activation measurements used to constrain astrophysical reaction rates, detector efficiency calibration, decay corrections in environmental and medical assays, and nuclear data. In this work, we revisit four tin nuclides: $^{110}$Sn (astrophysical p-process benchmarks), $^{113}$Sn (a calibration standard), $^{117\mathrm{m}}$Sn (nuclear medicine dosimetry), and $^{123\mathrm{m}}$Sn (nuclear data). All were produced via photon activation of natural tin and monitored by long-term HPGe spectrometry, providing an independent cross-check of evaluated decay data.

\subsection{$^{110}$Sn: Astrophysical Significance in the p-Process}

Precision half-life measurements for neutron-deficient isotopes near the $Z=50$, $N=50$ shell closures are essential for modeling the astrophysical p-process, which synthesizes rare proton-rich nuclei not produced via the s- or r-processes~\cite{Rauscher2001,Arnould2003}. The reaction network in this mass region is typically constrained by Hauser–Feshbach statistical model calculations, whose reliability hinges on inputs such as optical model potentials, level densities, and $\gamma$-strength functions~\cite{Goriely1998}. Because experimental cross sections at stellar energies (the Gamow window) are sparse, activation measurements and decay data—including half-lives—are used to normalize and validate rate predictions~\cite{Sauter1997,Bork1998,Chloupek1999,Gyurky2006}. For $^{110}$Sn, accurate half-life information improves the fidelity of cross-section measurements for reactions such as $^{106}$Cd$(\alpha,\gamma)^{110}$Sn, which probes the light p-nuclei production pathway up to $^{113}$In in the $Z=50$ region~\cite{Gyurky2006}. Independent half-life determinations using distinct production and counting methodologies therefore provide critical cross-checks on evaluated data and help reduce systematic uncertainties in nuclear astrophysics.

\subsection{$^{113}$Sn: Calibration and Standards}

The ground state of $^{113}$Sn decays by electron capture to $^{113}$In, emitting well-characterized $\gamma$ rays widely used for HPGe spectrometer calibration and tracer studies. Its half-life has been measured repeatedly over decades. Early systematic programs at Oak Ridge National Laboratory employed NaI(Tl), proportional counters, and ionization chambers, establishing values near $\sim$115~d~\cite{Emery1972}. Independent $\gamma$-spectrometric measurements with radiochemically purified samples corroborated this range~\cite{Lagoutine1972}. Large inter-comparison efforts coordinated by the International Atomic Energy Agency (IAEA) remeasured numerous $\gamma$ emitters (including $^{113}$Sn) with calibrated $4\pi$ ionization chambers, yielding high-precision decay constants~\cite{Houtermans1980}. Subsequent re-examinations at national standards laboratories incorporated improved detector stability, impurity analyses, and reference-source corrections~\cite{Hoppes1982,Unterweger1992}. These studies support the currently recommended value $T_{1/2}=\SI{115.09\pm0.03}{d}$ in the Nuclear Data Sheets (NDS) evaluation for $A=113$~\cite{Blachot2005}. Because $^{113}$Sn underlies detector calibration and inter-laboratory consistency, new independent determinations using distinct production routes and high-resolution gamma-ray spectrometry remain valuable.

\subsection{$^{117\mathrm{m}}$Sn: Nuclear Medicine and Dosimetry}

The isomeric state $^{117\mathrm{m}}$Sn (J$^{\pi}=11/2^{-}$) combines a monoenergetic 159 keV photon suitable for Single-Photon Emission Computed Tomography (SPECT) imaging with 127--155 keV conversion electrons for localized therapy. Its decay data are comprehensively evaluated in the NDS for A = 117~\cite{Blachot2002}, and its medical applications have been explored in modern radiopharmaceutical reviews~\cite{Boschi2019}. Its $\sim$14-day half-life enables distribution and radiopharmaceutical synthesis, including SnCl$_2$ and DOTA-based chelates for bone metastasis treatment. Traditional production routes employ neutron or proton reactions on tin and antimony, often yielding low specific activity or complex target chemistry~\cite{Maslov2011,Stevenson2015}. Photonuclear production via $^{118}$Sn($\gamma$,n)$^{117\mathrm{m}}$Sn using modern electron accelerators offers a clean and scalable alternative~\cite{Utsunomiya2011}.
Accurate half-life data for $^{117\mathrm{m}}$Sn are critical because the NDS recommends $T_{1/2}=\SI{13.76\pm0.04}{d}$~\cite{Blachot2002}, and even small deviations from this value can propagate into decay-correction protocols, patient dose planning, and quantitative imaging. Such discrepancies directly affect absorbed dose estimates and therapeutic efficacy, making high-precision measurements essential for reliable nuclear medicine applications.

\subsection{$^{123\mathrm{m}}$Sn: Nuclear Data Evaluations}

The isomeric state $^{123\mathrm{m}}$Sn (J$^{\pi}=3/2^{+}$) has a recommended half-life of $T_{1/2}=\SI{40.06(2)}{min}$ according to the NDS evaluation for A = 123~\cite{Chen2021}. This isomer plays a role in nuclear data libraries because its decay characteristics are used in activation analysis, decay heat calculations, and modeling of photonuclear reaction chains. Even small deviations from the evaluated half-life can propagate into decay-correction factors and impact the accuracy of cross-section measurements and applied nuclear science calculations. 

\section{Experimental Method}

\subsection{Activation}
\begin{figure*}[htbp]
    \centering
    \includegraphics[width=0.22\textwidth]{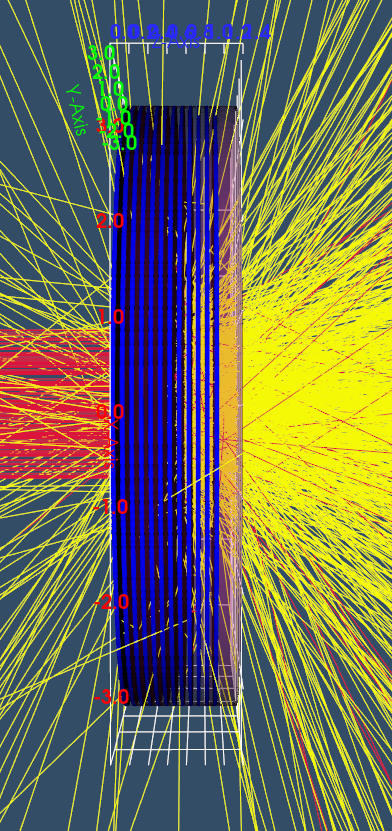}
    \caption{PHITS~\cite{PHITS2024} simulation of the irradiation setup. The plate-based tantalum converter (blue disks) is cooled by water flowing between the plates (transparent). The tin target is positioned just beyond the converter window (not shown). Yellow tracks represent bremsstrahlung photons, while red tracks correspond to electrons traveling from left to right.}
    \label{fig:setup}
\end{figure*}

A \SI{0.6186}{g} high‑purity natural tin target, composed of \SI{99.3}{\percent} Sn with a \SI{0.7}{\percent} Cu impurity fraction, was irradiated downstream of a \SI{6}{mm}-thick tantalum bremsstrahlung converter. The irradiation was performed using a TT‑300HE rhodotron delivering \SI{40}{MeV} electrons at a beam current of \SI{1}{\micro\ampere}, with a nominal beam spot diameter of \SI{20}{mm}. The converter consists of Ta plates, with water channels flowing between the plates as well as in front and back of the first and last plates. The irradiation geometry and particle transport are illustrated in Figure~\ref{fig:setup}, where bremsstrahlung photons (yellow tracks) and electrons (red tracks) are shown interacting with the converter and target.

\begin{figure*}[htbp]
    \centering
    \includegraphics[width=1.0\textwidth]{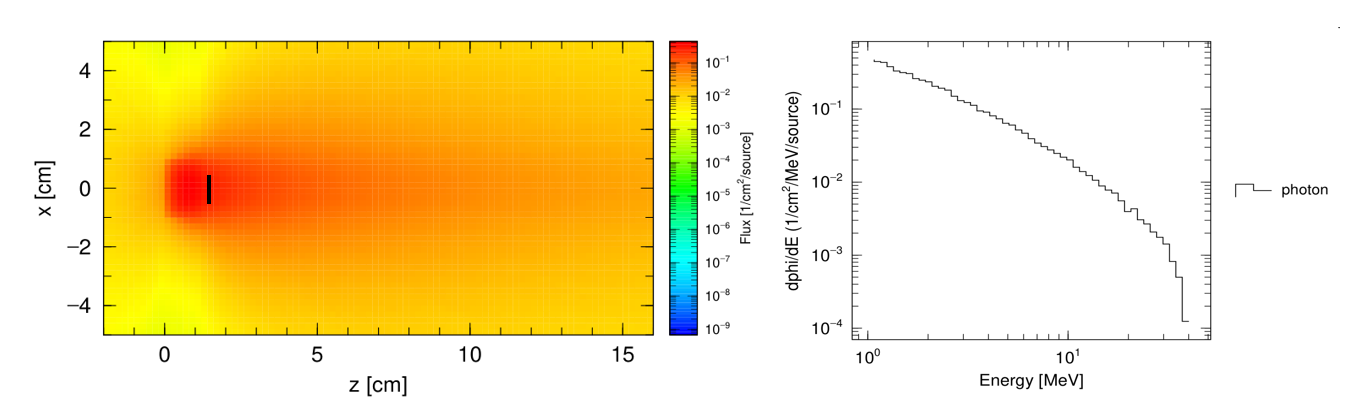}
    \caption{
    PHITS~\cite{PHITS2024} simulation results for a \SI{40}{MeV} electron beam (directed from left to right along the $z$‑axis) with a \SI{10}{mm} diameter spot. 
    The left panel shows the normalized total photon flux per incident electron. 
    The first tantalum converter plate, oriented perpendicular to the beam direction, is located at $z=\SI{0}{cm}$. 
    Absolute photon flux can be obtained by multiplying the color scale by the electron rate corresponding to \SI{1}{\micro\ampere} (\SI{6.24e12}{e^{-}\per\second}). 
    The black rectangle denotes the location of the tin target. 
    The right panel shows the bremsstrahlung photon energy flux at the mid‑plane of the target, per incident source electron.}
    \label{fig:flux}
\end{figure*}

The normalized photon flux distribution at the target location is shown in Figure~\ref{fig:flux}. This total energy-integrated photon flux map corresponds to a bremsstrahlung photon spectrum with \SI{40}{MeV} end-point energy. To obtain the absolute flux, the color scale must be multiplied by the electron rate for \SI{1}{\micro\ampere} beam current (\SI{6.24e12}{e^{-}/s}). The black square in Figure~\ref{fig:flux} indicates the tin target position. The irradiation time was \SI{10}{s} to minimize the buildup of potential interferences and dead-time during the counting of short-lived nuclides such as $^{123\mathrm{m}}$Sn, while producing sufficient $^{113}$Sn and $^{117\mathrm{m}}$Sn activities for long-term counting.

In addition, a second irradiation was performed as part of a broader test to measure cross sections at lower beam energy. For this, a tin target (\SI{0.9526}{g}) was placed behind a single \SI{3}{mm} tungsten plate (non-cooled) and irradiated under identical conditions of beam current (\SI{1}{\micro\ampere}) and irradiation time (\SI{20}{s}), but with an electron end-point energy of \SI{36}{MeV}. After activation, the target was moved for extended counting with \SI{180}{s} live-time intervals over more than one day to enable a precise determination of the $^{110}$Sn half-life. Only the $^{110}$Sn data originate from this lower-energy irradiation; all other isotopes ($^{113}$Sn, $^{117\mathrm{m}}$Sn, and $^{123\mathrm{m}}$Sn) were measured from the first irradiation using the \SI{6}{mm} Ta converter at \SI{40}{MeV}.

\subsection{$\gamma$-Spectral Processing and Calibration}
All spectra were parsed to extract run start times, live and real time, and channel-wise counts. For each spectrum, candidate peaks corresponding to the selected $\gamma$ lines ($E_i$) were located and fitted.

\paragraph{Energy, Resolution and Efficiency Calibrations}
To ensure accurate identification of $\gamma$-ray lines and assess detector performance, energy, resolution, and efficiency calibrations were performed using standard reference peaks from multiple calibration sources.

\begin{figure}[htbp]
    \centering
    \includegraphics[width=0.48\textwidth]{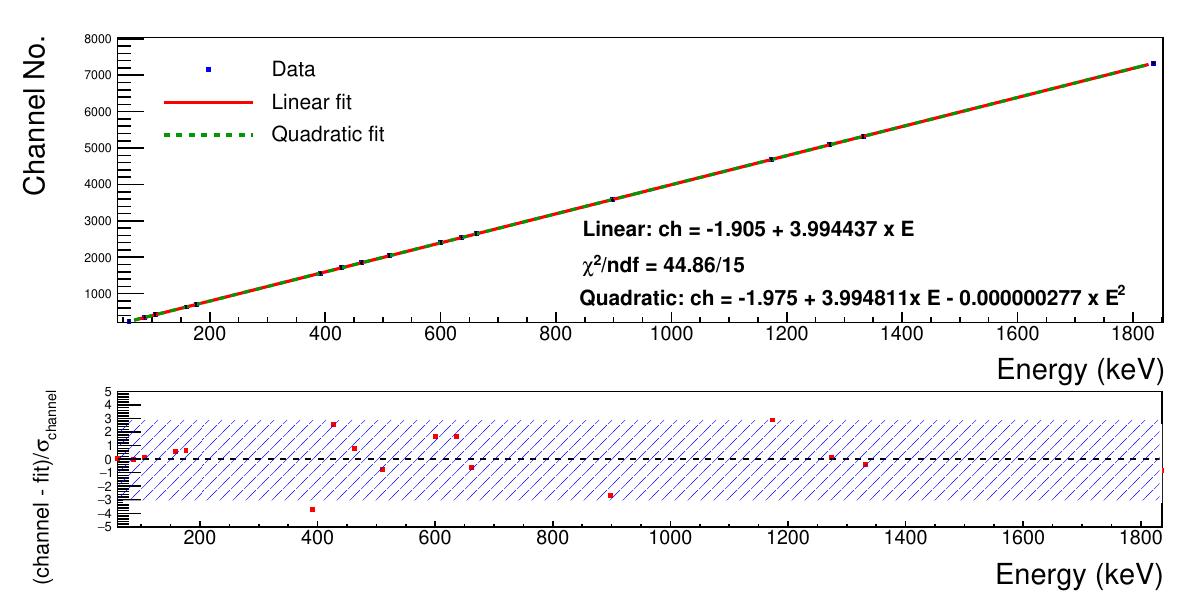}
    \caption{Energy calibration curve obtained by fitting channel centroids (blue markers; error bars are present but too small to be visible) to known $\gamma$-ray energies from multiple calibration sources. The linear fit (red line) shows excellent agreement, and the residuals in the lower panel confirm that a first-order polynomial is sufficient for the energy scale. The blue shaded region represents the $\pm 3\sigma$ confidence band for the residuals. A quadratic fit (dashed green line) was also tested; however, its $E^2$ coefficient ($-2.773\times10^{-7}\pm1.349\times10^{-6}$) is negligible and comparable to its uncertainty, so it was not adopted.}
    \label{fig:energy_calibration}
\end{figure}

\subparagraph{Energy Calibration}
The energy calibration was established by fitting known $(E_i,\mu_i)$ pairs, where $E_i$ denotes the $\gamma$ energy and $\mu_i$ the corresponding channel centroid. The calibration function is expressed as:
\begin{equation}
\mu = m_\mu \, E + c_\mu,
\label{eq:energy_cal}
\end{equation}
where $m_\mu = \SI{3.99444(72)}{\text{channel}/\keV}$ and $c_\mu = \SI{-1.905(383)}{\text{channel}}$ represent the gain and offset, respectively. The linear model was found to be adequate, as confirmed by the residuals shown in Fig.~\ref{fig:energy_calibration}. This calibration was used solely for peak identification and verification of assignments; all quantitative analyses (e.g., peak area extraction) were performed in the original channel domain without transforming spectra into energy space.

\begin{figure}[htbp]
    \centering
    \includegraphics[width=0.48\textwidth]{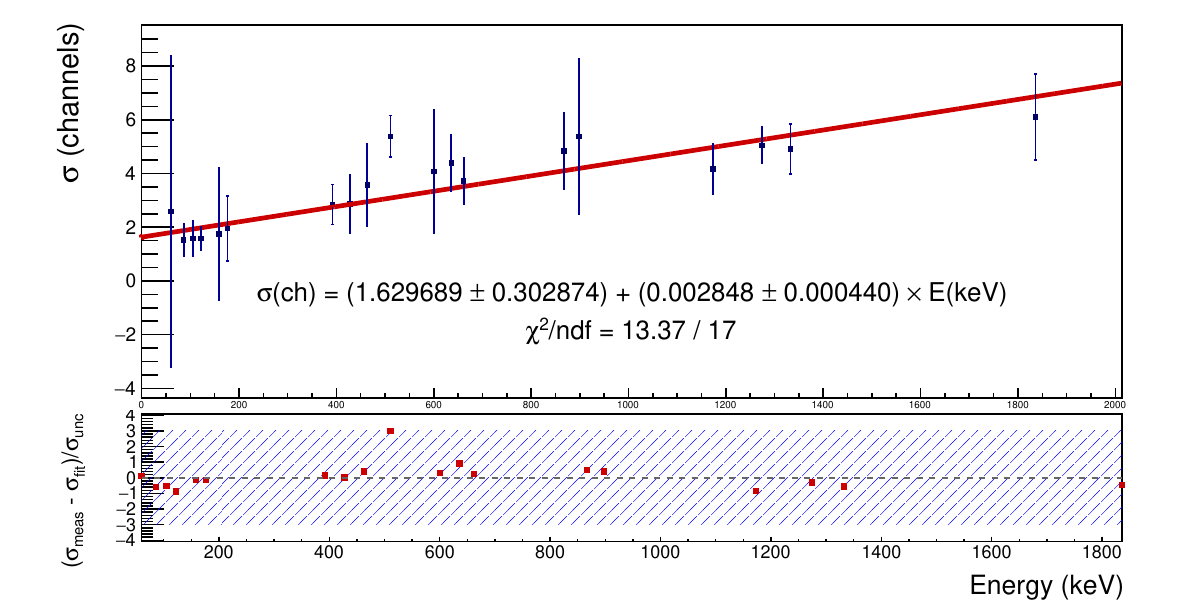}
    \caption{Resolution calibration curve showing the measured standard deviation, $\sigma$, as a function of $\gamma$ energy.}
    \label{fig:resolution_calibration}
\end{figure}

\subparagraph{Resolution Calibration}
The detector resolution, expressed as the Full Width at Half Maximum (FWHM), was derived from the fitted Gaussian standard deviation $\sigma$ using $\text{FWHM} = 2.355 \,\sigma$. The dependence of $\sigma$ on energy was found to be linear:
\begin{equation}
\sigma(\mu) = a + b \, E,
\label{eq:sigma_cal}
\end{equation}
where $a = \SI{1.6297(303)}{\text{channel}}$ and $b = \SI{0.002848(44)}{\text{channel}/\keV}$. This trend reflects contributions from electronic noise (constant term) and statistical fluctuations (energy-dependent term). The calibration curve and fitted function are shown in Figure~\ref{fig:resolution_calibration}, which also confirms the adequacy of the linear model across the energy range considered.

\subparagraph{Efficiency calibration}

Although absolute efficiency is not required for the half-life determination, which relies solely on
relative activity ratios, an efficiency calibration was nevertheless performed to ensure completeness
and internal consistency in all activity calculations.

\begin{figure}[htbp]
    \centering
    \includegraphics[width=0.48\textwidth]{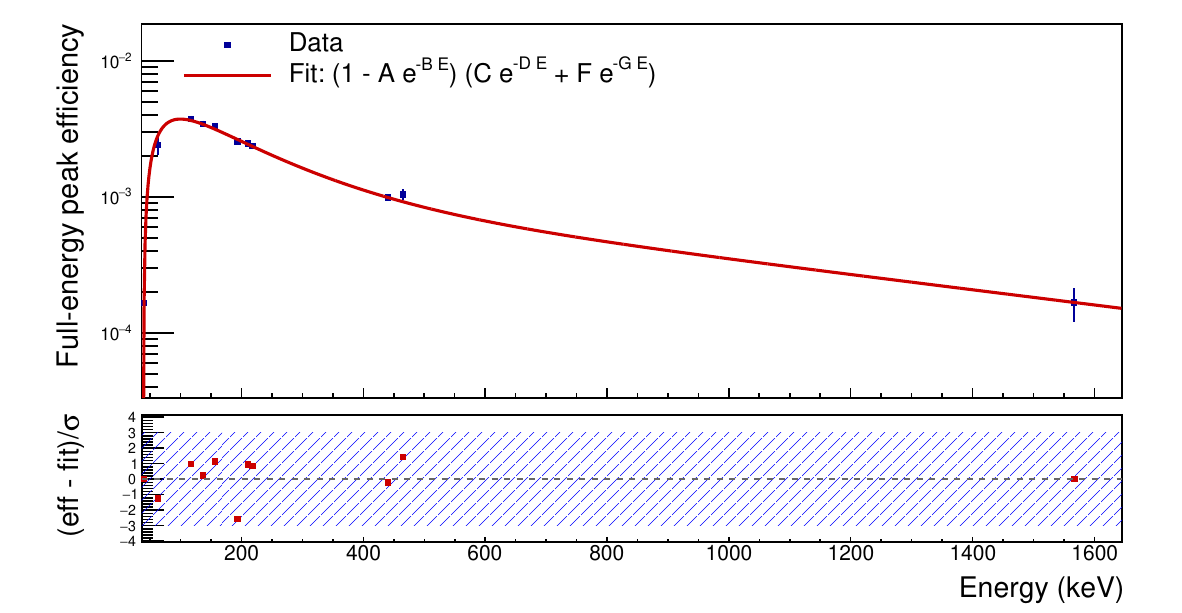}
    \caption{Efficiency calibration curve obtained by fitting the empirical function to measured 
    full-energy peak efficiencies from a 
    $^{229}$Th disk source. The top panel shows the data with the fitted function, while the 
    bottom panel presents standardized residuals with a shaded $\pm 3\sigma$ band. 
    The fit quality is characterized by a chi-square of $\chi^{2} = 14.374$ with 
    $\mathrm{NDF} = 5$.}
    \label{fig:efficiency_calibration}
\end{figure}

The calibration measurements were acquired using a $^{229}$Th calibration disk source that had the same radius as the activated targets and was placed at the same location of target counting. As shown in Fig.~\ref{fig:efficiency_calibration}, all
calibration points lie within
$\pm 3\sigma$ of the fitted curve, confirming the consistent detector response across the energy range.

The full-energy peak efficiency was modeled using the empirical function:
\begin{equation}
\begin{aligned}
\epsilon(E) 
= \bigl(1 - A\,e^{-B\,E}\bigr)\,\Bigl(C\,e^{-D\,E} + F\,e^{-G\,E}\Bigr),
\end{aligned}
\label{eq:efficiency}
\end{equation}
where $E$ is the photon energy in keV. The best-fit parameters are the following:
\[
\begin{aligned}
A &= 3.259722 \pm  0.389687\,, \\[6pt]
B &= 3.013335\times10^{-2} \pm 3.018147\times10^{-3}\ \text{[keV}^{-1}]\,, \\[6pt]
C &= 1.246222\times10^{-3} \pm 2.971436\times10^{-4}\,, \\[6pt]
D &= 1.279700\times10^{-3} \pm 2.979005\times10^{-4}\ \text{[keV}^{-1}]\,, \\[6pt]
F &= 6.943779\times10^{-3} \pm 6.426730\times10^{-4}\,, \\[6pt]
G &= 7.275769\times10^{-3} \pm 8.709378\times10^{-4}\ \text{[keV}^{-1}].
\end{aligned}
\]

\subsection{Overview of Spectral Evolution}
\begin{figure*}[htbp]
    \centering
    \includegraphics[width=0.7\textwidth]{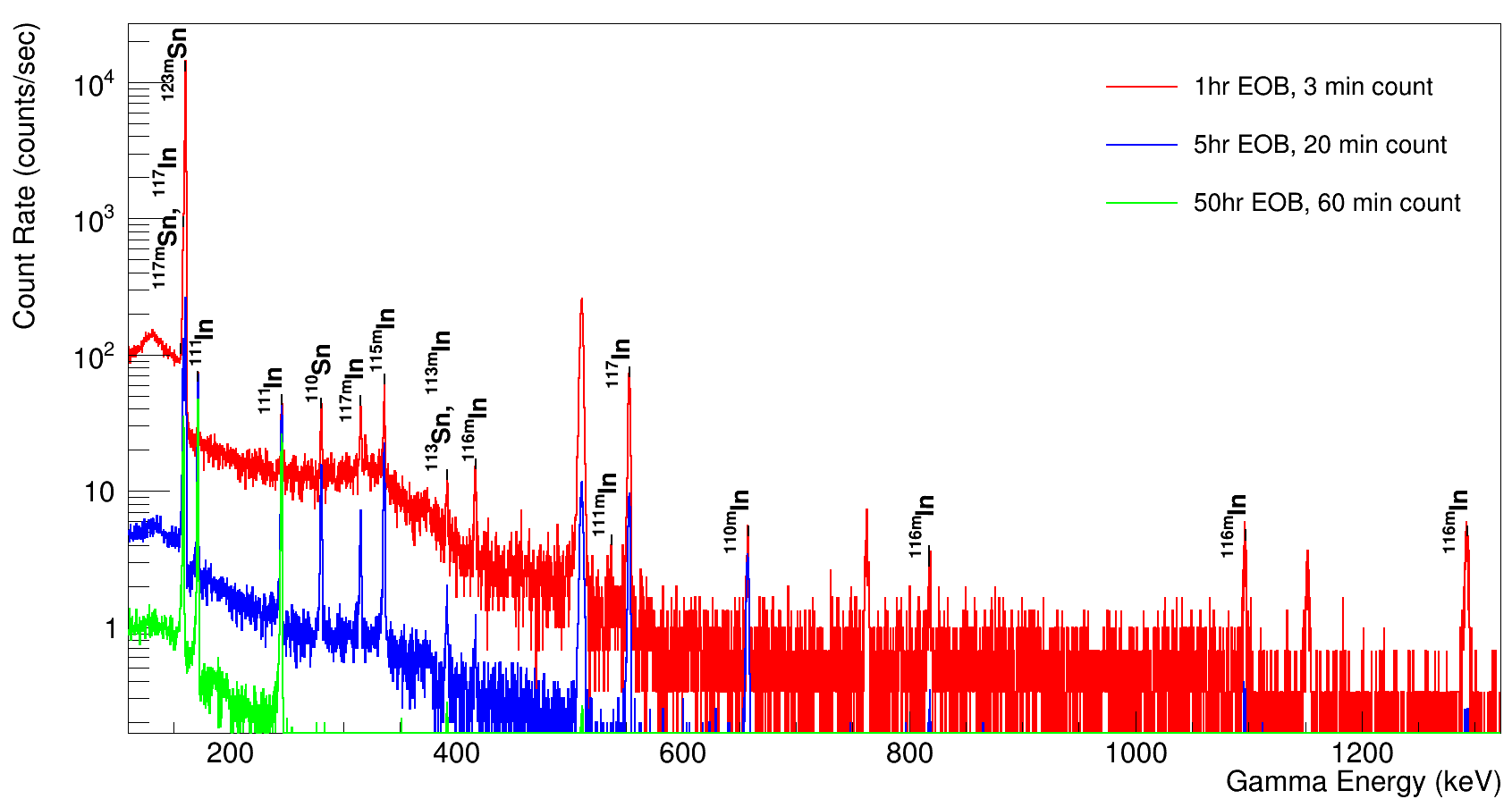}
    \caption{Full $\gamma$ spectra collected at 1~hour (red), 5~hours (blue), and 50~hours (green) post-irradiation. Counting times were 3, 20, and 60~minutes, respectively. Isotopes contributing major peaks are labeled.}
    \label{fig:unzoomed}
\end{figure*}

Figure~\ref{fig:unzoomed} shows three full $\gamma$ spectra collected at different times after End Of Bombardment (EOB): 1~hour (red), 5~hours (blue), and 50~hours (green). Counting times were 3~minutes, 20~minutes, and 60~minutes, respectively. The evolution illustrates the decay of short-lived isotopes and the persistence of longer-lived species such as $^{117\mathrm{m}}$Sn and $^{113}$Sn. Major isotopes are labeled on the plot. 

\subsection{Isotope-Specific Peak Models}

\paragraph{$^{110}$Sn (\SI{280.49}{keV}): Gaussian with two-component tail and piecewise background}
The $^{110}$Sn line exhibited low-energy tailing and a background slope change across the centroid. We modeled it as a mixture of two Exponentially Modified Gaussians (EMGs) sharing the same Gaussian width, plus a piecewise linear background:
\begin{equation}
\begin{aligned}
f_{110}(E) &=
\eta\,\mathrm{EMG}(E;\,C,\mu,\sigma,\tau_1)
+ (1-\eta)\,\mathrm{EMG}(E;\,C,\mu,\sigma,\tau_2) \\
&\quad + 
\begin{cases}
b_1\,E + m_1, & E < \mu,\\
b_2\,E + m_2, & E \ge \mu,
\end{cases}
\end{aligned}
\label{eq:gauss_tail}
\end{equation}
where $\eta$ is the mixture weight, $C$ is the net area under the peak with $\mu$ channels as the peak centroid, $\sigma$ the peak width, $\tau_{1,2}$ the tail parameters, and $b,c$ the background terms. Centroid, width, tail parameters, and $\eta$ were fixed.

The EMG function is defined as the convolution of a Gaussian with an exponential decay:
\begin{equation}
\begin{aligned}
\mathrm{EMG}(E;\,C,\mu,\sigma,\tau) &=
\frac{C}{2\tau}\,
\exp\!\left(\frac{E-\mu}{\tau} + \frac{\sigma^2}{2\tau^2}\right) \\
&\qquad \times
\mathrm{erfc}\!\left(\frac{1}{\sqrt{2}}\left(\frac{E-\mu}{\sigma} + \frac{\sigma}{\tau}\right)\right),
\end{aligned}
\label{eq:emg_def}
\end{equation}
The complementary error function $\mathrm{erfc}(x)$ accounts for the asymmetric tailing toward lower energies.

\begin{figure*}[htbp]
    \centering
    \includegraphics[width=0.8\textwidth]{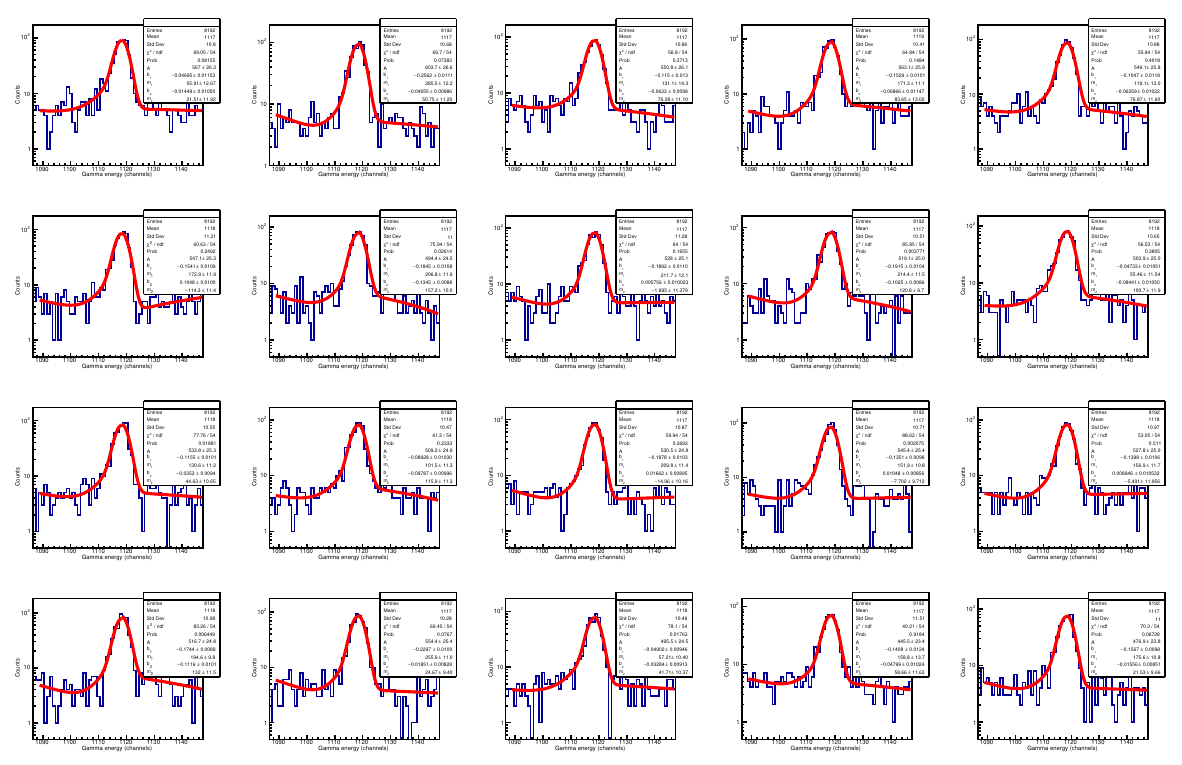}
    \caption{Twenty sequential $\gamma$-ray spectra for $^{110}$Sn, recorded between 6~h~8~min and 7~h~14~min post-irradiation. The panels are arranged to reflect the temporal evolution of the spectra, progressing from the top-left to the bottom-right across rows. The dominant peak at channel 1118.38 corresponds to the \SI{280.49}{keV} transition of $^{110}$Sn. Each spectrum was acquired over a 180~s live counting interval. The detector dead-time for the fitted data is below 0.8\%. Spectra were fitted using a Gaussian function with a two-component tail and a piecewise background model. Detailed fitting parameters are visible when zooming into the PDF version.}
    \label{fig:sn110_fit}
\end{figure*}

Figure~\ref{fig:sn110_fit} presents 20 sequential spectra fitted using the tail-corrected Gaussian model. The panels are arranged such that the sequence progresses from the top-left panel to the bottom-right, moving across each row from left to right. This layout reflects the temporal evolution of the collected spectra, with each subsequent panel corresponding to a later acquisition time. The dominant peak at channel $\mu = 1118.38 \pm 0.11$ represents the \SI{280.49(11)}{\keV} transition of $^{110}$Sn.

\paragraph{$^{113}$Sn (\SI{391.697}{keV}): single-Gaussian model}
The $^{113}$Sn ground-state transition emits a photon with energy \SI{391.697}{keV} and was well isolated and fitted with a single Gaussian plus linear background:
\begin{equation}
f_{113}(E) = \frac{C}{\sigma\sqrt{2\pi}}\;\exp\!\left[-\frac{(E-\mu)^2}{2\sigma^2}\right] + b\,E + c,
\label{eq:energy_fit}
\end{equation}
where $C$ is the net area under the peak, $\sigma$ the peak width, and $b,c$ the background terms.

\begin{figure*}[htbp]
    \centering
    \includegraphics[width=0.8\textwidth]{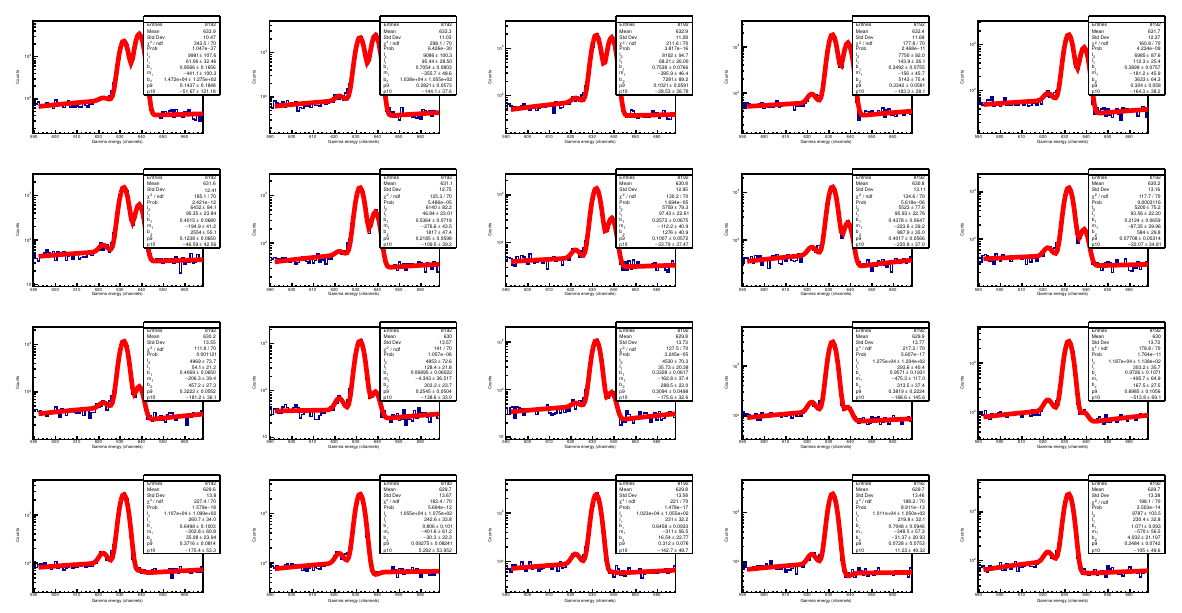}
    \caption{Sequential spectra for $^{117\mathrm{m}}$Sn between 5--15.5 hours (top-left to bottom-right panels) post-irradiation, fitted with a triple-Gaussian model. The main peak at channel (peak centroid) 631.663 corresponds to \SI{158.56}{keV} ($^{117\mathrm{m}}$Sn), flanked by peaks at channels 621.955 (\SI{156.2}{keV}) and 638.784 (\SI{160.34}{keV}), from $^{117\mathrm{m}}$Sn and $^{123\mathrm{m}}$Sn, respectively.}
    \label{fig:sn117m_fit}
\end{figure*}

\paragraph{$^{117\mathrm{m}}$Sn (\SI{158.56}{keV}), and $^{123\mathrm{m}}$Sn (\SI{160.2}{keV}): triple-Gaussian with piecewise linear background}
Figure~\ref{fig:sn117m_fit} shows sequential spectra collected between 5 and 15.5 hours post-irradiation, illustrating the triple-Gaussian fit applied to the \SI{158.56}{keV} region. The middle peak corresponds to $^{117\mathrm{m}}$Sn (\SI{158.56}{keV}, intensity 86.4\%, $T_{1/2}\approx$14~d), while the rightmost peak originates from $^{123\mathrm{m}}$Sn (\SI{160.34}{keV}, intensity 85.69\%, $T_{1/2}\approx$40.06~min). The lower energy peak located at peak centroid channel number 621.955 (\SI{156.2}{keV}, intensity 2.11\% from $^{117\mathrm{m}}$Sn)  contributes a minor but noticeable component in the spectra. 

As shown in figure~\ref{fig:sn117m_fit}, the $^{117\mathrm{m}}$Sn, and $^{123\mathrm{m}}$Sn region exhibits the three adjacent peaks. To de-convolve these, we used a triple-Gaussian with a piecewise linear background that changes slope at the main centroid $\mu_2$, corresponding to the middle peak (\SI{158.56}{keV}):
\begin{strip}
\begin{equation}
\begin{aligned}
f_{117\text{m}}(E) &=
\begin{cases}
\displaystyle
\frac{C_2}{\sigma_2\sqrt{2\pi}}\,
    e^{-\frac{(E-\mu_2)^2}{2\sigma_2^2}}
+
\frac{C_1}{\sigma_2\sqrt{2\pi}}\,
    e^{-\frac{(E-\mu_1)^2}{2\sigma_2^2}}
+
b_1\,E + m_1,
&
E < \mu_2,
\\[1.2em]
\displaystyle
\frac{C_2}{\sigma_2\sqrt{2\pi}}\,
    e^{-\frac{(E-\mu_2)^2}{2\sigma_2^2}}
+
\frac{C_3}{\sigma_2\sqrt{2\pi}}\,
    e^{-\frac{(E-\mu_3)^2}{2\sigma_2^2}}
+
b_2\,E + m_2,
&
E \ge \mu_2.
\end{cases}
\end{aligned}
\label{eq:triple_gauss}
\end{equation}
\end{strip}
where $C_{1,2,3}$ are the net areas of the lower-adjacent, middle, and upper-adjacent peaks, respectively; $\mu_{1,2,3}$ their centroids; and $\sigma_2$ the common width. Centroids and width were fixed to calibration values; areas and background terms were floated.

Because $C_{3}$ represents the net area under the $^{123\mathrm{m}}$Sn's \SI{160.34}{keV} peak, the $^{123\mathrm{m}}$Sn half-life analysis is performed using the same function in equation~\ref{eq:triple_gauss}.

\subsection{Iterative Decay Correction and Half-Life Refinement}
For each spectrum, the activity at the reference time was computed from the fitted peak net area using:
\begin{equation}
\begin{aligned}
A(t) &=
\frac{C_i\,\lambda\,\dfrac{\Delta t_{real}}{\Delta t_{live}}}{\epsilon_i\,I_\gamma \;\big(1 - e^{-\lambda\,\Delta t_{real}}\big)},
\end{aligned}
\label{eq:activity}
\end{equation}
where $\epsilon_i$ is the efficiency (applied though not required for half-life determination), $I_\gamma$ the emission probability, $\Delta t_{real}$ the real counting time, $\Delta t_{live}$ the live counting time, and $\lambda=\ln 2/T_{1/2}$ the decay constant.

We note that the formulation used in Eq.~\eqref{eq:activity} is fully consistent with established practice in activation-based $\gamma$-spectrometry measurements, as demonstrated in peer-reviewed studies in closely related experimental contexts~\cite{Zaman2018,Naik2020}. The factor $\Delta t_{real}/\Delta t_{live}$ corrects for fractional counting losses within the acquisition window, while the bracketed term $(1 - e^{-\lambda \Delta t_{real}})$ accounts for true physical decay during the finite real elapsed interval. For our multichannel analyzer (MCA) acquisition system operating under low dead-time conditions ($<0.2\%$ for $^{117\mathrm{m}}$Sn and $<0.8\%$ for $^{110}$Sn during fitted intervals), dead time is non-extending and stable throughout each counting interval ($\Delta t \ll T_{1/2}$). We evaluated alternative formulations (such as replacing $\Delta t_{real}$ in the decay correction factor with $\Delta t_{live}$, or implementing a generalized dead-time algorithm); the resulting differences in the extracted half-lives were strictly below $0.01\%$, demonstrating that the choice of correction formulation has a negligible effect on our results.

Half-life determination followed an iterative procedure:
\begin{enumerate}
    \item Initialize $T_{1/2}$ with the NDS recommended value, $T_{1/2}^{(0)}$.
    \item Correct activities using Eq.~\eqref{eq:activity}.
    \item Fit activity vs. time to:
    \begin{equation}
    A(t) = A_0\,e^{-\lambda^{(n)} t}.
    \label{eq:decaylaw}
    \end{equation}
    \item Update $T_{1/2}^{(n)}$, the n$^{th}$ iteration value of the half-life, and repeat until:
    \[
    \left|\frac{T_{1/2}^{(n)} - T_{1/2}^{(n-1)}}{T_{1/2}^{(n-1)}}\right| < 1\%.
    \]
\end{enumerate}

\section{Results}

\subsection{$^{110}$Sn Decay Analysis}
\begin{figure}[htbp]
    \centering
    \includegraphics[width=0.48\textwidth]{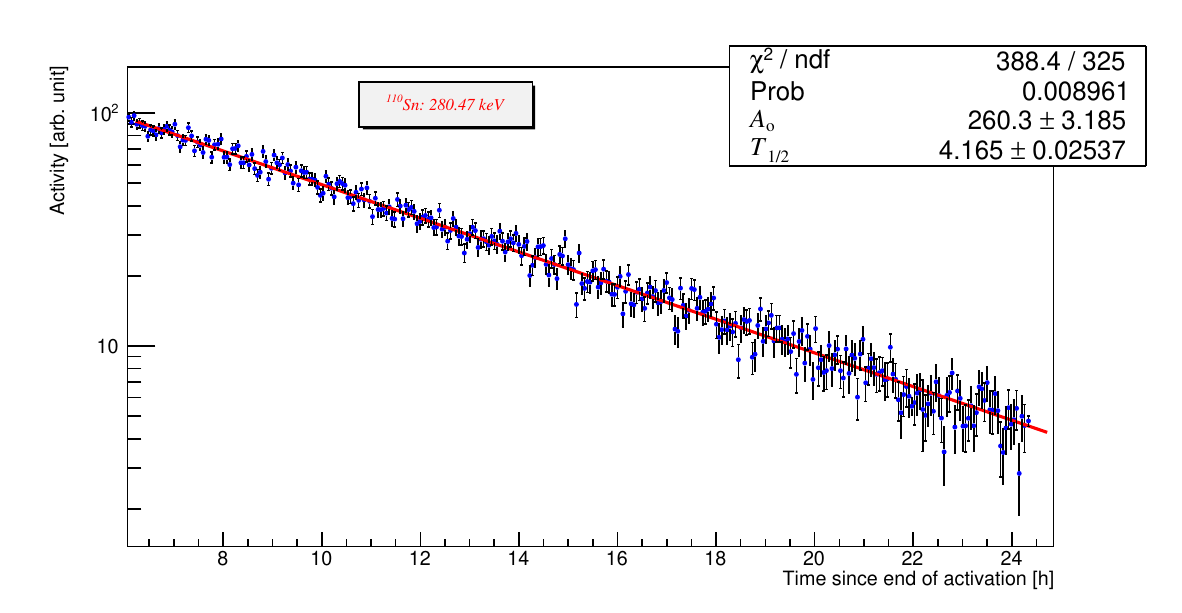}
    \caption{Decay curve for $^{110}$Sn derived from the \SI{280.49}{keV} $\gamma$ line. The data were fitted with a single-exponential decay model (red line) to extract the half-life. Early measurements (0–6~h post-irradiation) were excluded due to high dead-time (higher than 10\%), while all fitted points have dead-time between 0.30\% and 0.77\%. Live counting time was set to \SI{180}{s}.}
    \label{fig:sn110}
\end{figure}

Figure~\ref{fig:sn110} illustrates the decay behavior of $^{110}$Sn derived from the \SI{280.49}{keV} $\gamma$ line. The fit quality is characterized by \(\chi^2/\mathrm{ndf}=388.4/325\) and \(p=8.96\times10^{-3}\), indicating a statistically acceptable model. The extracted half-life is:
\[
T_{1/2}\big(^{110}\mathrm{Sn}\ \mathrm{g.s.}\big) = \SI{4.165(25)}{h},
\]
which is in close agreement with the NDS recommended value of \SI{4.154(4)}{h}~\cite{Gurdal2012}. 

\subsection{$^{113}$Sn Decay Analysis}
\begin{figure}[htbp]
    \centering
    \includegraphics[width=0.48\textwidth]{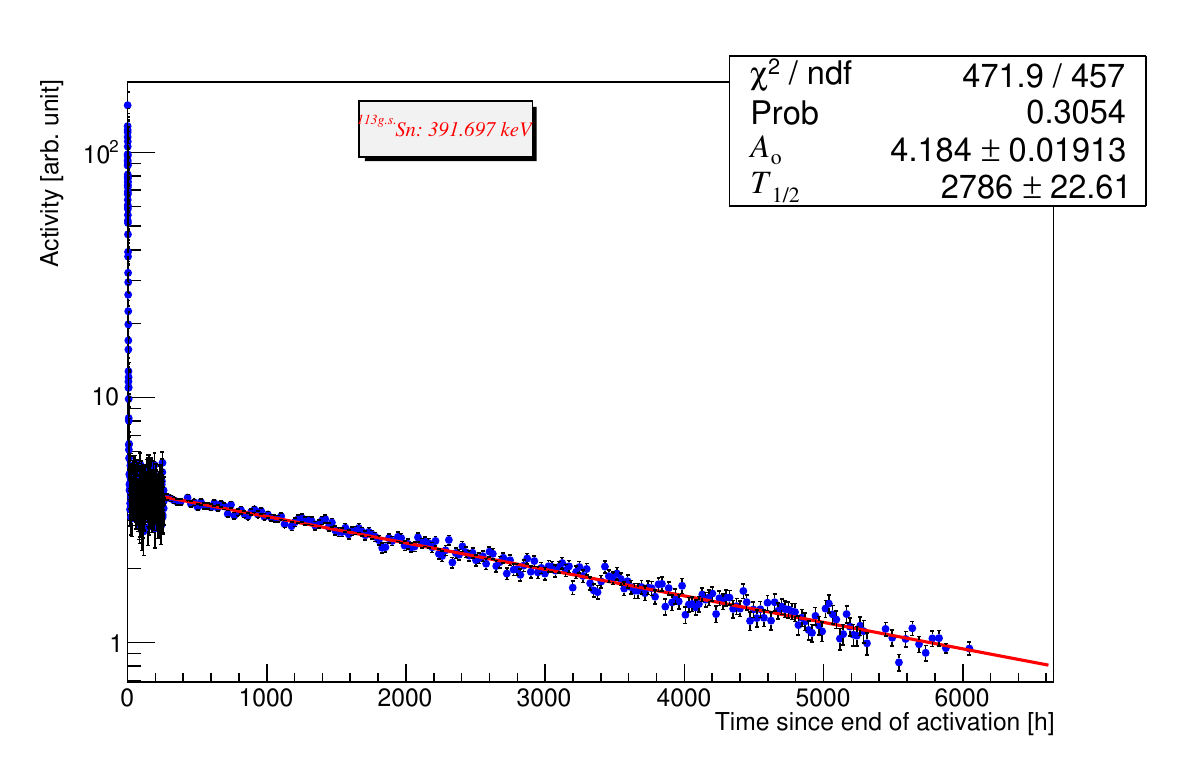}
    \caption{Decay curve for $^{113}$Sn based on the \SI{391.697}{keV} $\gamma$ line. Data points represent measured activity over time; the solid line shows the best-fit single-exponential decay. The fit includes data starting at 15~h (900~min) after the end of irradiation to avoid contamination from the \SI{391.698}{keV} line of $^{113}$In ($T_{1/2} = 99.476$~min) and extends to slightly beyond 6000~h post-irradiation. The counting live time was adjusted throughout the measurements to appropriately allow for other isotopic half-life determinations.}
    \label{fig:sn113}
\end{figure}

Figure~\ref{fig:sn113} shows the activity-versus-time dataset and the single-exponential fit for the \SI{391.697}{keV} line of $^{113}$Sn. The fit quality is evident (\(\chi^2/\mathrm{ndf}=471.9/457\), \(p=0.305\)), indicating that a single-component exponential decay adequately describes the data over the acquisition period.

From the fit to Eq.~\eqref{eq:decaylaw}, we obtain:
\begin{align}
T_{1/2}\big(^{113}\mathrm{Sn}\ \mathrm{g.s.}\big) = \SI{116.08 \pm 0.94}{d}.
\label{eq:result}
\end{align}
The relative deviation is:
\[
\frac{116.08-115.09}{115.09}\times100\%\approx 0.86\%.
\]

To verify that the \SI{391.698}{keV} emission from isomeric $^{113\mathrm{m}}$In ($T_{1/2} = \SI{99.476}{min}$), produced directly during photon activation, does not distort the fitted $^{113}$Sn decay slope, a sensitivity test was conducted by systematically varying the fitting start time between \SI{5}{h} and \SI{30}{h} post-irradiation. While fitting initialized before \SI{10}{h} exhibited a slight deviation due to transient non-equilibrium of $^{113\mathrm{m}}$In, the extracted half-life remained stable to within $0.05\%$ for all start times $\ge \SI{15}{h}$. Restricting the fit to $t \ge \SI{15}{h}$ (corresponding to $>9$ half-lives of $^{113\mathrm{m}}$In) ensures full secular equilibrium and confirms that direct precursor contributions do not bias the reported $^{113}$Sn half-life.

\subsection{$^{117\mathrm{m}}$Sn Decay Analysis}
\begin{figure}[htbp]
    \centering
    \includegraphics[width=0.48\textwidth]{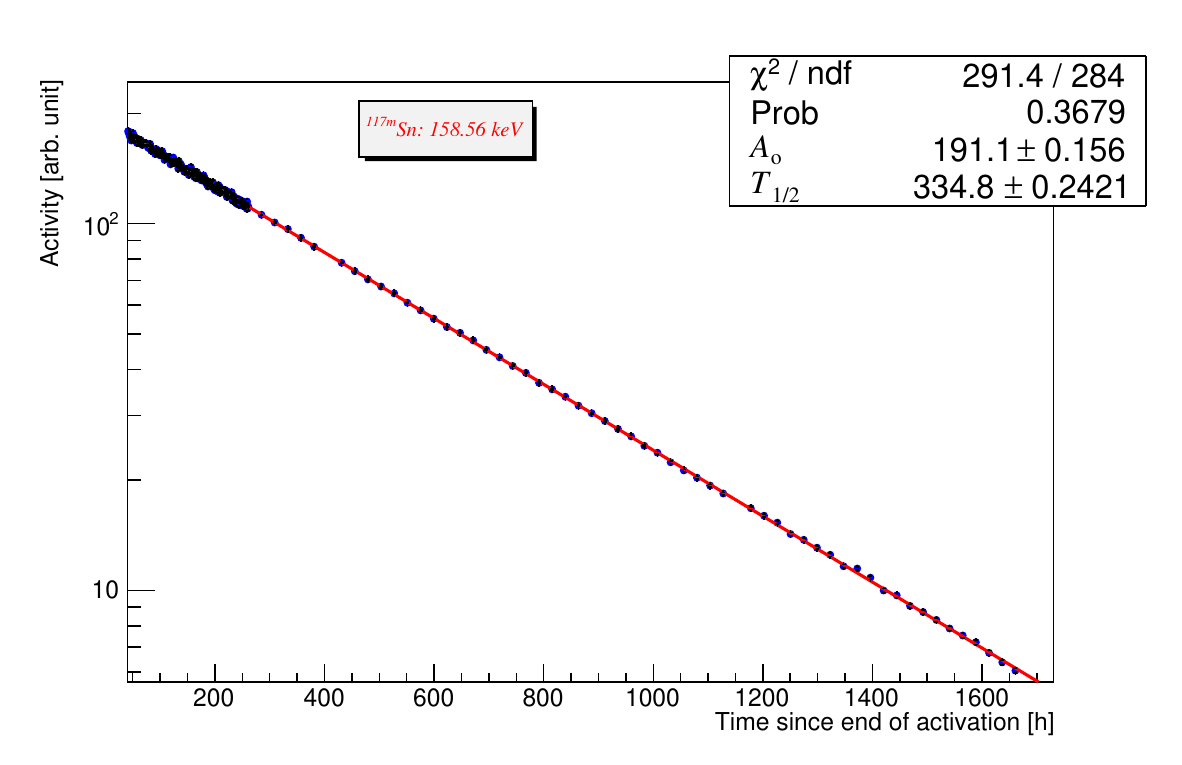}
    \caption{Decay curve for $^{117\mathrm{m}}$Sn based on the \SI{158.56}{keV} $\gamma$ line. Measured activity points are fitted with a single-exponential decay model. To eliminate early feeding from the decay of $^{117}$In (both ground state, $T_{1/2} = 43.2$~min, and isomeric state, $T_{1/2} = 116.1$~min), the fit was restricted to data collected starting 50~h after the end of irradiation. Additionally, the dead-time of the counting spectra that are used in this fit was always less than 0.2\%.}
    \label{fig:sn117m}
\end{figure}

Figure~\ref{fig:sn117m} presents the decay analysis for $^{117\mathrm{m}}$Sn using the \SI{158.56}{keV} line. The fit quality is evident (\(\chi^2/\mathrm{ndf}=312.4/305\), \(p=0.389\)), and the extracted half-life is:
\[
T_{1/2}\big(^{117\mathrm{m}}\mathrm{Sn}\big) = \SI{13.95(1)}{d}.
\]
Compared to the NDS recommended value of \SI{13.76\pm0.04}{d}~\cite{Blachot2002}, the deviation is small but statistically significant. Such differences can propagate into decay-correction protocols and dosimetric calculations in nuclear medicine, directly impacting absorbed dose estimates and therapeutic efficacy.

\subsection{$^{123\mathrm{m}}$Sn Decay Analysis}
\begin{figure}[htbp]
    \centering
    \includegraphics[width=0.48\textwidth]{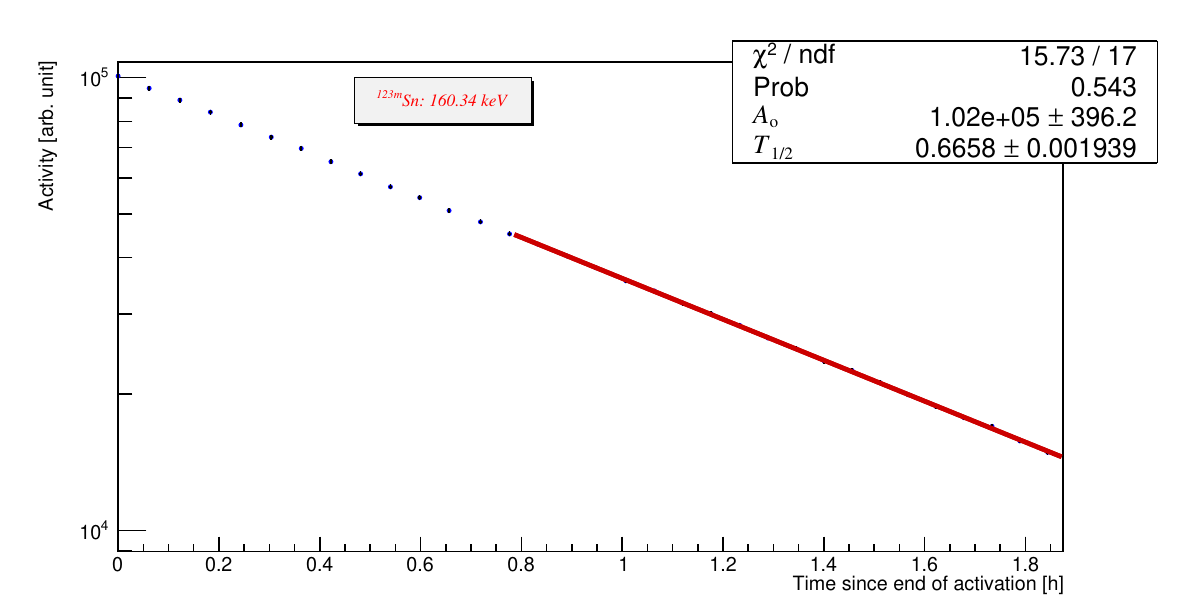}
    \caption{Decay curve for $^{123\mathrm{m}}$Sn. Activity data points were fitted with a single-exponential decay model (red line) using an iterative correction approach. Only measurements with dead-time below 10\% were included in the fit after applying the dead-time correction, see equation~\ref{eq:activity}. Each point represents \SI{180}{s} of live counting time.}
    \label{fig:sn123m}
\end{figure}

Figure~\ref{fig:sn123m} shows the decay analysis for $^{123\mathrm{m}}$Sn based on its characteristic \SI{160.34}{keV} $\gamma$ emission. The fit was performed following the same iterative correction approach, yielding \(\chi^2/\mathrm{ndf}=15.73/17\) and \(p=0.543\). The extracted half-life is:
\[
T_{1/2}\big(^{123\mathrm{m}}\mathrm{Sn}\big) = \SI{39.95(12)}{min}.
\]

Compared to the NDS recommended value of \SI{40.06(2)}{min}~\cite{Chen2021}, the difference is small. 

\subsection{Summary of Results}
Table~\ref{tab:halflives} summarizes the measured half-lives for $^{110}$Sn, $^{113}$Sn, $^{117\mathrm{m}}$Sn and $^{123\mathrm{m}}$Sn, obtained using isotope-specific peak models and iterative decay correction, and compares them to the NDS recommended values. Deviations between the measured and evaluated half-lives were quantified using a z-score (standardized difference), defined as
\[
z = \frac{x_{\mathrm{meas}} - x_{\mathrm{eval}}}{\sqrt{\sigma_{\mathrm{meas}}^{2} + \sigma_{\mathrm{eval}}^{2}}},
\]
where $\sigma_{\mathrm{meas}}$ and $\sigma_{\mathrm{eval}}$ represent the associated uncertainties. While most isotopes exhibit good agreement within uncertainties ($|z| \lesssim 1$), the case of $^{117\mathrm{m}}$Sn shows a significantly larger deviation, with $z = 4.608$. This elevated z-score clearly underscores the statistical significance of the discrepancy, indicating that the deviation is unlikely to arise from random uncertainty alone. Such a result highlights the need for closer examination, particularly given its potential implications for calibration, dosimetry, and astrophysical modeling.

\begin{table*}[htbp]
\centering
\caption{Measured half-lives compared to NDS recommended values. Uncertainties are $1\sigma$ statistical from the exponential fit. The last column shows the z-score value.}
\label{tab:halflives}
\begin{tabular}{lccccc}
\toprule
Isotope & Energy (keV) & Fit Model & This Work $T_{1/2}$ & NDS $T_{1/2}$ & z-score \\
\midrule
$^{110}$Sn & 280.49 & Gaussian + tail & \SI{4.165(25)}{h} & \SI{4.154(4)}{h}~\cite{Gurdal2012} & 0.435 \\
$^{113}$Sn & 391.697 & Single Gaussian & \SI{116.08(94)}{d} & \SI{115.09(3)}{d}~\cite{Blachot2005} & 1.053 \\
$^{117\mathrm{m}}$Sn & 158.56 & Triple Gaussian & \SI{13.95(1)}{d} & \SI{13.76(4)}{d}~\cite{Blachot2002} & 4.608 \\
$^{123\mathrm{m}}$Sn & 160.34 & Triple Gaussian & \SI{39.95(12)}{min} & \SI{40.06(2)}{min}~\cite{Chen2021} & -0.904 \\
\bottomrule
\end{tabular}
\end{table*}

\subsection{Uncertainty Considerations}

The uncertainties reported in Table~\ref{tab:halflives} represent $1\sigma$ statistical uncertainties obtained from the weighted exponential fits. Systematic uncertainties were quantitatively evaluated across several categories:
\begin{itemize}
    \item \textbf{Dead-time and pulse pile-up:} Spectra utilized for half-life determinations were restricted to low dead-time domains ($<0.8\%$ for $^{110}$Sn and $<0.2\%$ for $^{117\mathrm{m}}$Sn). Uncertainty in the live-time clock correction was assessed by comparing alternative correction algorithms and was found to contribute $<0.03\%$ systematic uncertainty to the decay constant.
    \item \textbf{Peak integration and background model:} Systematic effects from region-of-interest boundary selection, tailing parameters, and piecewise background slopes were evaluated by varying fitting intervals and background formulations across the decay sequence. The resulting area variation contributed $<0.05\%$ uncertainty to the extracted half-lives.
    \item \textbf{Geometry and detector stability:} Samples were mounted in rigid reproducibility holders ensuring position reproducibility within $\Delta x < \SI{0.1}{mm}$. Detector gain and resolution stability were monitored across runs using calibration lines; centroid drift remained within $<0.1$ channel, contributing $<0.02\%$ to the decay slope uncertainty.
    \item \textbf{Timing accuracy:} System acquisition start/stop times and hardware timestamps were verified against synchronized network time servers, yielding negligible systematic error ($<10^{-4}\%$).
\end{itemize}
For $^{117\mathrm{m}}$Sn, where the statistical uncertainty is exceptionally small ($\pm\SI{0.01}{d}$, or $0.072\%$), the combined systematic uncertainty in quadrature is estimated to be $\pm\SI{0.008}{d}$ ($0.057\%$). Because this systematic bound is smaller than the quoted statistical uncertainty, the reported precision is realistic and robust.

\section{Discussion}

This work provides independent verification of evaluated half-lives through photon activation of natural tin and long-term HPGe monitoring. In several cases, these measurements represent the only determinations obtained via a photo-nuclear production route for the isotope or its isomeric state, offering a unique complement to the established dataset.

Generally speaking, the NDS methodology consolidates results from a wide range of experimental techniques, including charged-particle reactions such as $(\alpha,2n)$, $(p,t)$, and $(d,p)$, Coulomb excitation, photo-nuclear processes, and neutron-capture routes. For example, the adopted half-life of $^{110}$Sn (4.154~h) reflects a weighted average of measurements from $(\alpha,2n\gamma)$ reaction on $^{108}$Cd, $(p,t)$ reaction on $^{112}$Sn, Coulomb excitation studies, and other routes, with uncertainties primarily statistical. Similarly, $^{113}$Sn values originate from activation via $(\alpha,n\gamma)$ reaction on $^{110}$Cd, $(p,n\gamma)$ reaction on $^{113}$In, $(p,t)$ reaction on $^{115}$Sn, and $(n,\gamma)$ reaction on $^{112}$Sn, complemented by decay studies following $\varepsilon$ capture. For $^{117\mathrm{m}}$Sn, the NDS recommendation (13.76~d) aggregates results from $\beta^-$ decay chains of $^{117}$In, internal transition studies, and multiple charged-particle reactions, such as $(\alpha,n\gamma)$ reaction on $^{114}$Cd, and $(d,p)$ reaction on $^{116}$Sn, and neutron radiative capture on $^{116}$Sn, while $^{123\mathrm{m}}$Sn values are based on decay monitoring after $(d,p)$ and $(n,\gamma)$ reactions on $^{122}$Sn.

Our approach differs in two key aspects: (i) the use of photon activation, which enables simultaneous production of multiple isotopes under identical conditions and minimizes contamination from competing reaction channels, and (ii) strict control of systematic effects, including dead-time correction and exclusion of early-time spectra to mitigate feeding from short-lived precursors (e.g., $^{117}$In). These measures allow high-precision determinations, defined here as results with relative uncertainties below 1\%, ensuring reliability for applications in nuclear medicine and astrophysical modeling.

Key findings for individual isotopes are summarized below:

\textbf{$^{110}$Sn}: Our result agrees within uncertainties with the NDS value, supporting its reliability for astrophysical $p$-process modeling, where accurate half-lives constrain reaction rates near the $Z=50$, $N=50$ shell closures.

\textbf{$^{113}$Sn}: Although the measured half-life differs by 0.86\% from the NDS recommendation (with z-score = 1.053), this discrepancy—, though small, highlights the importance of reexamining reference values used in HPGe efficiency calibrations. Even sub-percent deviations can propagate into systematic errors in detector standards.

\textbf{$^{117\mathrm{m}}$Sn}: The measured value of \SI{13.95(1)}{d} exceeds the older NDS recommendation of \SI{13.76(4)}{d} by $1.38\%$ ($z = 4.608$), representing a statistically significant deviation. However, our result does not stand isolated from previous experimental work; it is in close agreement with the precision experimental measurement of \SI{13.91(3)}{d} reported by Domingo et al.~\cite{Domingo2017} using HPGe spectroscopy. The current NDS evaluation reflects a weighted average that incorporates historical measurements favoring shorter half-lives ($13.60$--$13.76$~d). By utilizing high-resolution HPGe spectroscopy on photon-activated tin and initiating fits after \SI{50}{h} to eliminate $^{117}$In feeding, our measurement provides an independent, low-background verification that corroborates the recent determination by Domingo et al.~\cite{Domingo2017}. Because $^{117\mathrm{m}}$Sn is an important therapeutic radionuclide, adopting an accurate half-life is critical for absorbed dose calculations and clinical radiopharmaceutical release criteria.

\textbf{$^{123\mathrm{m}}$Sn}: The measured half-life is about 0.28\% lower than the NDS recommendation (z-score = -0.904), within one standard deviation. 

Overall, these results underscore the importance of independent verification using alternative production mechanisms and stringent uncertainty criteria. Periodic reevaluation of medically relevant isotopes remains essential to maintain confidence in nuclear data libraries and their downstream applications.

\section{Conclusions}
Two short photon activations of natural tin were performed to produce measurable activities of $^{110}$Sn, $^{113}$Sn, $^{117\mathrm{m}}$Sn, and $^{123\mathrm{m}}$Sn suitable for short- and long-term HPGe monitoring. The primary irradiation used a \SI{6}{mm} Ta converter and a \SI{40}{MeV} electron beam, while a secondary irradiation at \SI{36}{MeV} with a single \SI{3}{mm} W plate was carried out as part of a broader photo-nuclear reaction cross-section test. Extended counting intervals (\SI{180}{s}) were applied only to the second irradiation to enable a precise determination of the $^{110}$Sn half-life; all other isotopes were analyzed from the first irradiation.

Based on isotope-specific peak models (single Gaussian, triple Gaussian, and Gaussian with tail) and iterative decay correction, the following half-lives were determined:
\begin{itemize}
    \item $^{110}$Sn: \(T_{1/2}=\SI{4.165(25)}{h}\),
    \item $^{113}$Sn: \(T_{1/2}=\SI{116.08(94)}{d}\),
    \item $^{117\mathrm{m}}$Sn: \(T_{1/2}=\SI{13.95(1)}{d}\),
    \item $^{123\mathrm{m}}$Sn: \(T_{1/2}=\SI{39.95(12)}{min}\).
\end{itemize}
The measured half-lives of $^{110}$Sn, $^{113}$Sn, and $^{123\mathrm{m}}$Sn agree with NDS recommendations within experimental uncertainties ($|z| \le 1.05$). For $^{117\mathrm{m}}$Sn, our measured value of \SI{13.95(1)}{d} deviates significantly from the older NDS recommendation (\SI{13.76(4)}{d}, $z = 4.608$), but closely aligns with the experimental measurement of Domingo et al.~\cite{Domingo2017} (\SI{13.91(3)}{d}). These results provide valuable independent experimental input for nuclear data libraries, demonstrating the capability of photon activation to produce clean decay benchmarks for astrophysics, detector calibration, and medical dosimetry.

\section*{Acknowledgments}
The authors would like to thank the accelerator operators and the health physics team at NorthStar Medical Radioisotopes (NMR) for the stable and safe operation of the machine and safety and logistical support.

%% ---------------- Bibliography ----------------
\bibliographystyle{elsarticle-harv}
\bibliography{refs}

\end{document}